\documentclass[a4paper,12pt]{article}
\usepackage{graphicx}
\usepackage{amssymb,amsmath}
\usepackage{textcomp}
\usepackage{subfigure}
\usepackage{feynmp}
\usepackage{cite}
\title{Scaling Properties of Firearms Homicides in Brazilian cities}
\author{Airton Deppman}
\date{Instituto de F\'isica \\  Universidade de S\~ao Paulo - Brazil \\ email: deppman@if.usp.br}

\begin{document}

\maketitle

\begin{abstract}
 The recent quantitative approaches for studing several aspects of urban life and infrastructure have shown that scale properties allow to understand many features of urban infrastructure and of human activity in cities. In this work, an analysis based on the complexity of cities network is performed for data on Brazilian cities firearms homicides. Due to the diversity in Brazilian population and cities, this is an interesting test for  theories  recently proposed. The superlinear power-law behavior for number of homicides as a function of city population, with exponent $\beta=1.15$ is obtained. The hypothesis that fractal structures can be formed in cities as well as in larger networks is tested, indicating that indeed self-similarity may be found in networks connecting several cities.
\end{abstract}


\section{Introduction}

The formation and growth of cities have being a long-standing concern, but only with the advent of large database providing information on social aspects of human life is that more rigorous scientific approaches to the social sciences have been made possible. While methods based on historical aspects of cities transformation were important to establish the most general aspects of urban development~\cite{Mumford}, only recently more quantitative analyses did indicate new directions in which our knowledge about cities can develop~\cite{Batty, WestCityNature, Netto}. Although still mostly phenomenological, new models have appeared that explain aspects of socioeconomic activity and human interactions in more solid grounds, allowing for the discovery of patterns that, before the recent developments, were unsuspected of~\cite{Bettencourt1, Bettencourt2, Bettencourt3} and triggering the emergence of a {\it science for cities} propitiated by  accurate information about how people live and interact, what is made possible by the advances in information technology. The scientific approach to understand cities organic evolution and how people live in the urban environment produced some promising results, revealing patterns that apparently are valid across the globe. These patterns are emergent characteristics of  cities~\cite{Batty2,WestScale} since they were not planned in any way, and therefore reflects fundamental aspects of human activity in urban centers that, in a coarse grained analysis, are determinant to shape urban structures and the way people interact in the urban environment.

One fascinating finding is the fact that many aspects of urban life and city infrastructure increase as a power-law function of the city size, measured by the size of its population, that is
\begin{equation}
  N=N_o \nu^{\beta}\, \label{powerlaw}
\end{equation}
where $N$ is the quantity of some mensurable feature of the city, as number of gas station, number of patents produced, number of squares in the city, while $\nu$ is the city population, with $N_o$ and $\beta$ being constants to be determined from data analysis, the first one representing a reference value for the quantity measured and the second one being the slope of the logarithm increase of the measured quantity $N$. It turns out that the exponent $\beta$ is in many cases different from unit, a result that shows that the usual {\it per capita}, or linear, method os comparison of socioeconomic or infrastructural data between different cities or countries is not necessarily the best form of analysis. Most interesting, studies have shown that when $N$ refers to an infrastructure quantity, the exponent is $\beta<1$, resulting in a sublinear increase of infrastructure with city size, while when $N$ refers to a measure of human activity, such as wages or crimes, the increase is superlinear, with $\beta>1$. Surprisingly, the exponents are around the values $\beta=0.85$ in one case and $\beta=1.15$ in the other, independently of other characteristics such as culture, ethnicity or country where the city is placed, while $N_o$ may vary from one place to the other. A consequence of these findings is that the increase of city size leads to more efficient usage of its infrastructure, while the pace of human activity is accelerated~\cite{Chowell,BattyFractalCities, Schlapfer}. Among the socioeconomic aspects of urban living, crimes is one of those that have deserved considerable attention~\cite{crimes1,crimes2}. The power-law behavior of urban related quatities stems from scale invariant aspects of cities that have been associated to a fractal geometry of the cities\cite{BattyFractalCities}, what means that districts and neighborhoods display features that are similar to those found in the city but at a smaller scale, reflecting a property called self-similarity. There are evidences that the self-similarity can be observed at large as well, as for instance in the flux of cars and trucks along roads connecting several cities.

In this work we analyze firearms homicides in Brazilian cities under the light of the power-law theory. Brazil is one of the largest countries in the world, with a population larger than two hundred million people living in more than 5 thousand cities of different sizes, cultures and history. The country is formed by hundreds of different ethnic groups living mostly spread over the territory and mixed with the other ethnic groups. Also, about 85\% of its population lives in urban environment, with cities as large as S\~ao Paulo with more than ten million people. Brazilian cities, therefore, offer an interesting test for the theory of power-law growth of human activity with the size of cities. The number of homicides in the country is dramatically high, reaching to more than sixty thousand homicides each year, most of them being caused by firearms injuries. A more scientific approach to investigate how homicides are distributed across the country and how cities networks facilitate or obstacle the criminal activity can be of great help to the development of more effective security policies. The analysis allow us to verify if self-similarity can be observed at large scales encompassing several cities.

\section{Methodology}

The present study is based on data collected from the study Mapa da Viol\^encia, a systematic analysis of violence in Brazil as compared to Latin America and other countries. In 2016, a careful survey of firearms homicides in the period 2012-2014 was carried out, considering the numbers registered in 3083 cities, corresponding to about 55\% of the Brazilian cities. The information is collected from death certificates, which gives date and place of death, and its causes. A discussion about the quality of the data collected from death certificates is available in Refs.~\cite{Mello, Ramos} where, for instance, attention is called upon the fact that in the case of death due to injuries caused by firearms the
local of death registered may be different from that of the crime. The study presented in Ref~\cite{MV} informs if the death was caused by self inflicted, accidental or intentional injuries, but in the analysis performed here there is no attempt to segregate those causes. For a relatively smal number of cases, the cause of injury cannot be determined, but the estimate from Mapa da Viol\^encia is that around 95\% of death by firearms result from intentional injuries. Populational information is calculated based on census data extrapolated according to the methodology from DATASUS~\cite{datasus} from data   of the Brazilian population census carried out in 2010.

During the period covered by the survey, Brazilian law about the use of firearms was not significantly changed and have been very restrictive about the possibility of acquiring and carrying firearms since 2004. The effects of the introduction of more restrictive law was a reduction of about 5\% in the number of firearms deaths in the two following years, after which the number remained rather stable for another period of four years~\cite{MV}. In the period of the survey, from 2012 to 2014, the number of crimes was increasing each year at a rate approximately equal to that observed before the establishment of the restrictive law. The total number of deaths in each year of the survey are: 40.077 in 2012, 40.369 in 2013 and 42.291 in 2014, summing up to 122.737 for the three-year period. In 2017, the number of homicides in Brazil was already above sixty thousand.

The present analysis is based on the recent theory for cities which predicts a superlinear power-law behavior for the number of firearms homicides, according to Eq.~\ref{powerlaw}\cite{Batty2,WestCityNature,Chowell,BattyComplexity,Bettencourt4}. Since this is one of the negative aspects of human interaction in the urban environment, it is supposed to increase with an exponent $\beta \sim 1.15$. Previous studies have stressed the fact that the power-law analysis must be carried out for each country individually~\cite{WestScale}, since socioeconomic activity can change significantly from one country to the other due to cultural aspects, different legislations, among other aspects, although the reference values, $N_o$, can vary in a significant way from country to country, the power-law exponent with $\beta=1.15$ seems to be an universal feature.  In the present study we also perform the same analysis for each individual Brazilian state. The results allow us to understand the limitations of the theory, and how more specific analysis could be carried out.

\begin{figure}[h]
\center
\subfigure[national][Firearms homicides in Brazilian cities.]{\includegraphics[width=7.7cm]{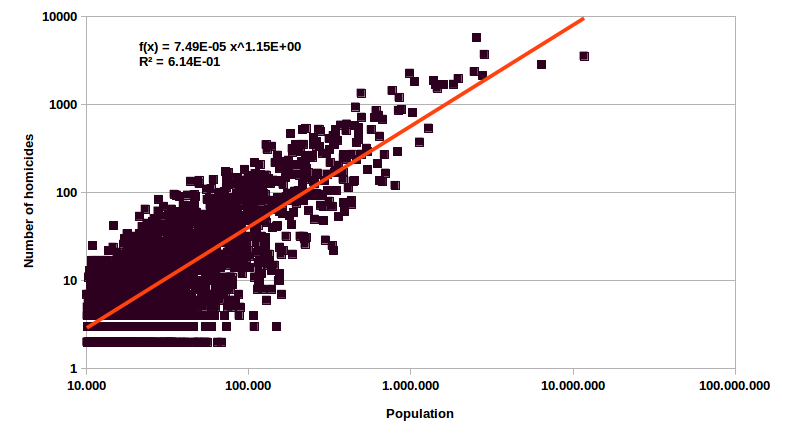}} 
\qquad
\subfigure[SP][Rank of the firearms homicide risk corrected by city scale.]{\includegraphics[width=7.7cm]{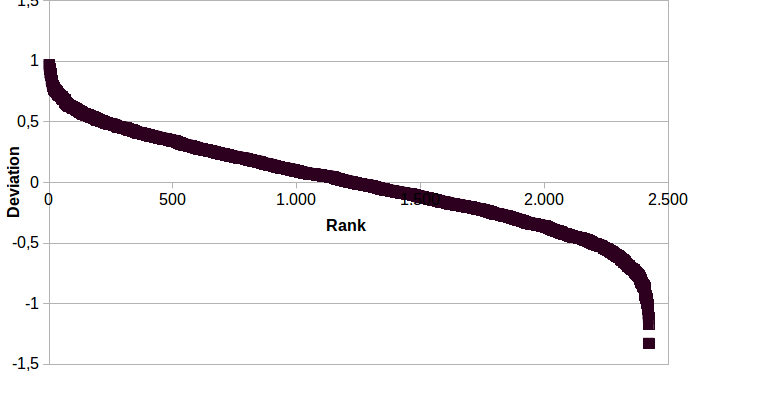}} \label{national}
\caption{(a) Scaling behavior of the number of firearms homicides in Brazilian cities. The trend line results to have slope $\beta=1,15$ $[1.08,1.27]$ (b) Deviation of number of homicides from the power-law prediction.} \label{national}
\end{figure}

The analysis performed here is based on the total number of firearms death for the period 2012-2014. Those numbers include intentional firearm deaths inflicted by motivations that can scarcely be associated to urban living, as  crimes of passion or suicide, for instance. These events may have strong effects on the analysis of small cities, where the total number of deaths in the period is small. In order to reduce effects of those events, we consider in the present analysis only cities with 2 or more firearm deaths in the period.

The parameter $\beta$ in Eq.~\ref{powerlaw} is determined by measuring the slope of the straight line in a log-log plot, representing the general behavior of the number of homicides as the size of cities increases.

\section{Results and Discussion}

The theoretical assumption underlying the power-law behavior given by Eq.~\ref{powerlaw} is that cities are constituted of networks through which information, goods and people flow. The interconnections between human interactions and city infrastructure, therefore, determine how people live in the cities and how cities grow shaped by human needs~\cite{WestScale, BattyComplexity}. Therefore cities, understood as its buildings, roads, squares and general infrastructure  but also the people living in it and their activities, constitute a complex adaptive system. In the theoretical framework used here, the power-law emerges as a consequence of scaling properties common to all cities in a given country~\cite{WestCityNature}. In this sense, Tokio is a scaled up version of Nagoia in the same way as New York is a scaled up version of Austin. However, one cannot compare Tokio to New York in the same lines, since cultural and historical aspects present in the development of one of them is not necessarily present in the development of the other.

The analysis of the overall behavior of the scaling properties of firearms homicides in Brazilian cities results to be in very good agreement with the theoretical predictions, as can be observed in Fig.~\ref{national}(a), where the number of firearms homicides for each city is plotted against the city population in a log-log plot. Despite the large fluctuation around the curve, we observe the data distributed along a straight line, unveiling the power-law behavior of the increase of crimes with the increase of the city size, measured by its population. The slope of the linear curve, which is equal to the exponent in the power law, results to be $\beta=1.15$, which is exactly the value expected from the general analysis of scaling properties in cities around the world~\cite{WestCityNature}. 

When analyzing the data under the light of the theoretical predictions, which implies in a superlinear increase of crimes with the city population, the rank of most dangerous cities is modified with respect to the usual rates of crimes per population city. Since the linear behavior is considered as a collective characteristics of the network of cities therefore the significant measure the rate of crimes to compare cities in the network must consider the natural evolution of criminality as the population size increases. The meaningful comparison is given by the residue, $R$, defined as
\begin{equation}
  R=log(N_H)-log(N_o x^{\beta})\,,
\end{equation}
where $N_H$ is the number of homicides observed in a city of the network characterized by the parameters $N_o$ and $\beta$ of the power-law behavior. The cities can be ranked by using the residue $R$ instead of the number of crimes {\it per capita}, as it is usually done.

The modification in the rank favors mostly the largest cities~\cite{Youn}, and this is observed in the case of the Brazilian data analyzed here. In Fig.~\ref{national}(b) we see the deviation $R$ plotted against the city rank ordered by decreasing deviation of the number of homicides observed from the value predicted by the power-law behavior. The curve is characteristic of this kind of analysis, and is similar to the corresponding one obtained in other works where different socioeconomic aspects were investigated~\cite{WestCityNature}. This result also confirms the theoretical predictions of universal behavior of human activity in cities. 

Both the linear behavior of homicides as a function of the population as the s-shape curve found for the rank of deviations from the linear behavior shows that the theory is confirmed for this socioeconomic variable in Brazilian cities. The confirmation of theoretical predictions obtained here is particularly relevant taking into account the diversity and size of the population, the complexity of its society and heterogeneity of territory, climate and history. In Brazil the different regions and cities received diverse cultural contributions along the few centuries of the country history. The urbanization process in Brazil was very fast: while only 31\% of its population was living in urban centers in 1940, about 85\% was living in cities in 2010. Considering the swift process of urbanization, one could expect that large differences between the cities could prevent the emergence of common behavior for the cities, thus it is a natural question to ask if the power-law behavior of cities is also observed in this country.
All these characteristics make the analysis of Brazilian data a critical test for a theory on cities that intends to present global aspects present in any city. 

\begin{figure}[t]
\center
\subfigure[Firearms homicides in Maranh\~ao.][Maranh\~ao]{\includegraphics[width=6.3cm]{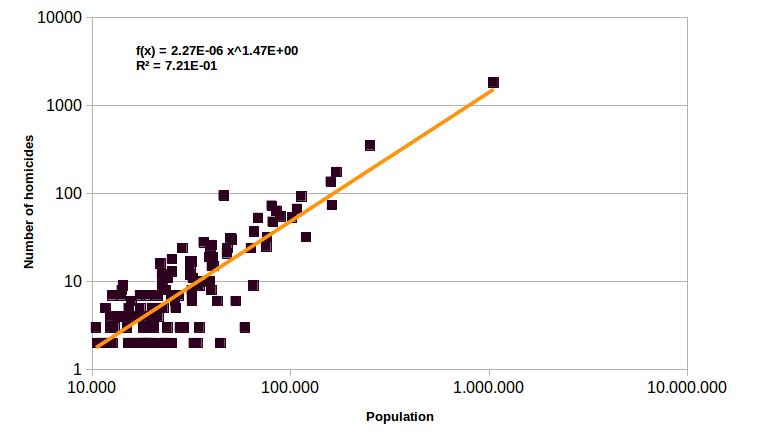}}
\qquad
\subfigure[Firearms homicides in S\~ao Paulo.][S\~ao Paulo]{\includegraphics[width=6.3cm]{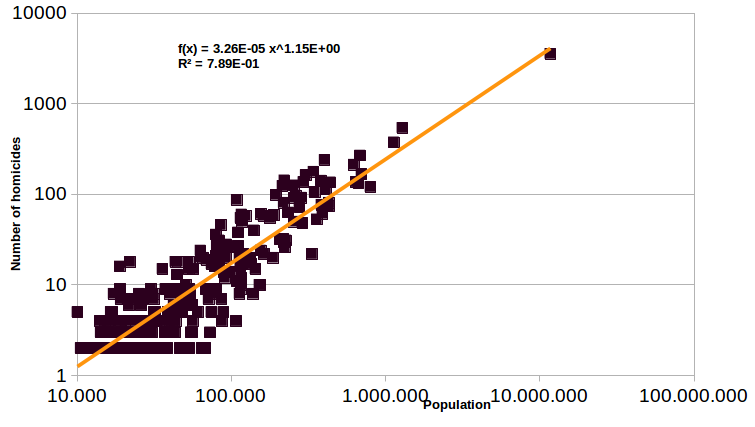}}
\qquad
\subfigure[Firearms homicides in Santa Catarina.][Santa Catarina]{\includegraphics[width=6.3cm]{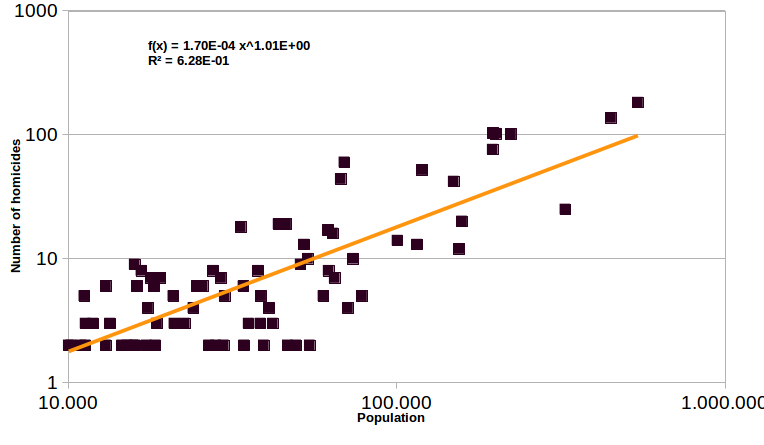}}
\caption{Scaling behavior of number of homicides by firearms in the states of Maranh\~ao, S\~ao Paulo and Santa Catarina. The number of homicides is plotted against the city population.} \label{states}
\end{figure}

Ranking cities according to its deviation from the power-law prediction shows some differences with respect to the traditional linear estimation, when {\it per capita} numbers are considered. In Table~\ref{tabrank} the ten most dangerous cities according to our analysis are displayed, and the new rank is compared to the {\it per capita}, or linear, rank. Although seven cities are among the ten most dangerous in both ranks, the order they appear is changed. But in some cases the deviation is larger: the 7th most dangerous city in the new rank was in position 22nd in the linear rank. All those cities have population in the range between 10.000 and 50.000, but on the other hand, cities with largest  populations are those that have their ranks shift to lower positions. In fact, our analysis show that cities like Rio de Janeiro, S\~ao Paulo, Bras\'ilia, Guarulhos and Campinas are those presenting largest downward shifts in the rank position.

\begin{table}[th]
\begin{tabular}{|c|c|c|}
\cline{1-3}
Rank & City                         & linear rank  \\ \cline{1-3}
1    &  Satuba - AL                 & 3            \\ \cline{1-3}
2    &  Murici - AL                 & 2            \\ \cline{1-3}
3    &  Conde - PB                  & 4            \\ \cline{1-3}
4    &  Mata de S\~ao Jo\~ao - BA   & 1            \\ \cline{1-3}
5    &  Pilar - AL                  & 6            \\ \cline{1-3}
6    &  Quixerer\'e - CE            & 12           \\ \cline{1-3}
7    &  Jaguaribara - CE            & 22           \\ \cline{1-3}
8    &  Eus\'ebio - CE              & 5            \\ \cline{1-3}
9    &  Pojuca - BA                 & 9            \\ \cline{1-3}
10   &  Marechal Deodoro - AL       & 11           \\ \cline{1-3}
\end{tabular}
\caption{Modification of city rank according to te power-law analysis, with respect to the linear rank.}  \label{tabrank}
\end{table} 

The results obtained are surprising in many aspects. Rio de Janeiro is commonly considered one of the most dangerous cities in Brazil, but according to this analysis, it is in fact a little safer than predicted for a typical Brazilian city with the same size. The same is valid for the city of S\~ao Paulo. These findings are of great importance for improving initiatives for reducing criminality. Resources should be directed to those cities with larger deviations from the linear behavior, and security policies should take this deviations to investigate how criminal activities spread over the territory and how they can be prevented. In this regard, analysis of regional crime rate, instead of the national one, can be of importance, and will be analyzed in the following.

\subsection{Robustness of superlinear power-law behavior}

So far we have shown that the theory of power-law behavior for firearms homicides with exponent around 1.15 is valid for Brazilian cities, despite several features that make this country different from many others across the globe. We showed that not only the general behavior of data is distributed as predicted, but also the deviation from the average behavior is in accordance with what is found in other analysis~\cite{MV}. The fact that the theoretical prediction is confirmed by the present analysis can also shed light on the mechanisms that allow the emergence of the complex behavior evidenced by the theory. Indeed, it shows that the power-law behavior and the values for the exponent $\beta=1.15$ is robust enough to be valid even in an environment where culture, geography and history varies considerably. The question that arises is: what are the relevant aspects for determining the power-law behavior observed, and the value of the exponent $\beta$?

Trying to answer such question, we analyze data for the Brazilian states separately. The hypothesis made here is that since the theoretical predictions are based on the scaling properties of the cities networks, and since states can be considered also as networks of cities, the self-similarity characteristic of fractals would induce the same power-law behavior of cities for each state. We analyzed data for 13 of the 26 Brazilian states, which are representative of the whole. We privileged those states with large population or larger number of firearms homicides in the period considered here. 

In. Fig.~\ref{states} we see the results for three of the states analyzed. We observe the linear curve representing the average behavior of cities within each state is found for all cases, what shows that the power-law behavior is valid. The exponents obtained by fitting the power-law distribution are shown in Table~\ref{tabbeta}, and we see that they are spread over a wide range, but in all cases they are above unit, confirming the theoretical prediction of superlinear behavior~\cite{WestCityNature}. For the states represented in Fig.~\ref{states} we have the following results: Maranh\~ao, which have the highest value for the exponent, with $\beta=1.47$; S\~ao Paulo, which has exactly the value expected by the theory, with $\beta=1.15$; and Santa Catarina, which presents the lowest homicide numbers and also the lowest exponent, with $\beta=1.005$, what means that the number of homicides increases almost linearly with population size for cities in that state. From the full set of analyzed data there is no clear relation between the reference number of homicides, $N_o$, for the states and their trend line slope. Similar variations in the slope of crime against city size was observed in India~\cite{Bettencourt-India}

At first sight, we observe that the superlinear power-law behavior is resilient to differences of culture, history and socioeconomic development, while the value for the exponent may vary considerably from one state to the other. One surprising result, however, is found for the state of S\~ao Paulo, for which $\beta=1.15$, what deserves further analysis. The discussion about the validity of the $\beta=1.15$ value for the exponent will be made below, but before that some practical remarks about the results of the present analysis will be made.

\begin{figure}[t]
 \centering
 \includegraphics[scale=0.6]{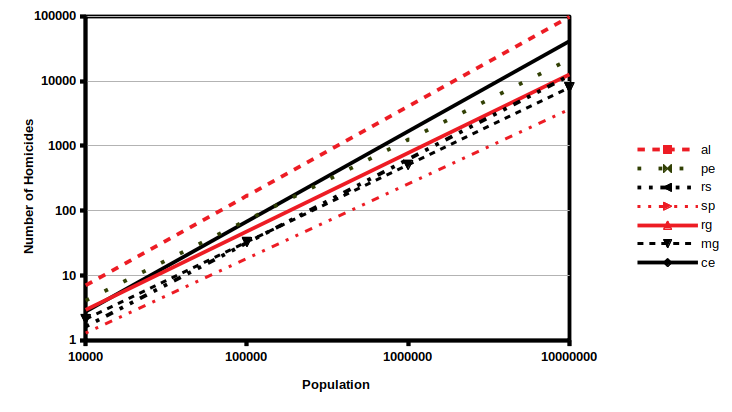}  
 \caption{Plot of equivalent-city number of firearms homicides for different states. The results for the states of Alagoas (al), Pernambuco (pe), Rio Grande do Sul (rs), S\~ao Paulo (sp), Rio de Janeiro (rj), Minas Gerais (mg) and Cear\'a (ce) are displayed. The results for S\~ao Paulo and Brazil present the same slope. }\label{estados}
\end{figure}

The new method of analysis show some unexpected results. Rio de Janeiro is usually mentioned as one of the most dangerous states in Brazil, a bad characteristic that is usually not shared by Paran\'a state. However, when analyzed under the light of the scaling properties, we observe that homicide by firearms in both states are very similar in both slope and reference values. In Rio de Janeiro, we have a smaller number of cities but they usually are larger than those present in Paran\'a state, where there are more cities with prevalence of small ones. This aspect explains why, when calculated by the usual {\it per capita} rates, Paran\'a state appears safer than Rio de Janeiro state. Also, the results for both sates are close to the average Brazilian result, showing that none of them can be placed among the most dangerous in the country when the power-law analysis is performed. This is an important result, since a large amount of resources are used to reduce violence in Rio de Janeiro, while from our analysis states like Alagoas, Cear\'a and Bahia are those that should deserve more governmental attention in order to reduce homicides rates.

\begin{table}[]
\begin{tabular}{|l|l|l|l}
\cline{1-3}
Network            & $\beta$ & range (95\% C.I.)&  \\ \cline{1-3}
Brazil             & 1.15    & {[}1.08,1,27{]} &  \\ \cline{1-3}
MA                 & 1,47    & {[}1.39,1.61{]} &  \\ \cline{1-3}
ES                 & 1.44    & {[}1.34,1.62{]} &  \\ \cline{1-3}
ES(Center-North)   & 1.35    & {[}1.22,1.37{]} &  \\ \cline{1-3}
ES(South)          & 1.32    & {[}1.21,1.39{]} &  \\ \cline{1-3}
ES(North)          & 1.18    & {[}1.15,1.20{]} &  \\ \cline{1-3}
SC                 & 1.005   & {[}0.925,1.104{]}& \\ \cline{1-3}
SP                 & 1.15    & {[}1.06, 1.127{]}& \\ \cline{1-3}
\end{tabular}
\caption{Slope obtained from the analysis  for a sample individual states and for separate networks in the state of Esp\'irito Santo.} \label{tabbeta}
\end{table}

The comparison of firearms homicide rates in the different states is made easier if we introduce the concept of equivalent-city, that is, if we calculate what would be the number of homicides predicted according to the theory if some hypothetical cities with different populations were found in those states. In Fig.~\ref{estados} we plot the crime rate per equivalent-city for different states as calculated by the power-law behavior obtained for the individual state analysis. The population of the equivalent-city ranges from ten thousand to ten million. Clearly, one will not find cities so large in many states, but this comparison of crimes in equivalent-cities shows what the rate of crimes would be if equivalent-cities could be found in each state. The plot allow us to identify the states with the most dangerous networks of cities, independently of their sizes. We observe that Alagoas is the most dangerous state in Brazil, and if a city with population of ten million could be found there, it would have around one hundred thousand homicides by firearms for the entire period of three years covered in this study, a number that can be compared to the real number observed for Brazil in the same period, which is 119.164. An interesting comparison is with the city of S\~ao Paulo, with a population of more than 11 million and where the number of homicides for the period was 3.568. Also, the effects of the different values for the exponent in the power-law behavior can be clearly seen by the different slopes. For some states, as it is the case for Rio Grande do Norte, Esp\'irito Santo and Maranh\~ao, small cities are safe for the Brazilian standards, but large cities are more dangerous than the average.

\subsection{Network effects on the power-law exponent}

The analysis of cities in a single state showed that two aspects of the theory are confirmed: the power-law behavior and the superlinear exponent. Now we will study the possible causes of the large differences among the exponents obtained. Our hypothesis is that the variation in the values for $\beta$ is related to the fact that cities of a single state are not comprised in a single network, while cities in an entire country may be considered as a single network.
The reason for this difference may be people circulation control at the international borders, what does not happen at borders between two national states. 
The first step is to understand why the cities in the state of S\~ao Paulo would be an exception the the rule, since for this state it was obtained $\beta=1.15$. S\~ao Paulo is the richest state in Brazil, is supplied by good roads and does not present relevant geographical obstacles to the flux of people, what allows for fast and easy connections between its cities, which is one of the main aspects for the power-law behavior of cities characteristics. Also, aside a few exception, S\~ao Paulo state has no touristic spots of national importance, and the touristic places usually receive much more people from S\~ao Paulo state than from other states or from other countries. S\~ao Paulo city is one exception to this, but the number of tourists is small compared to its huge population, therefore the effects of tourists in the number of homicides can be considered negligible. Taking these aspects into account we can regard the cities in S\~ao Paulo as a single network with much more connections among themselves than with cities from other states.

\begin{figure}[t]
\center
\subfigure[ES][Firearms homicides in Esp\'irito Santo state. Dots represent the number of homicides in different cities in the state: orange dots for those in the southern network and blak dots for those in the center-northern network. Orange circle represent the cities in the subnetwork in the northern region. Black dashed line represents the trend line for the entire state, resulting in $\beta=1.44$. Black continuous line represents the trend line for the center-north network region of the state. Orange continuous line represents the trend line for the south region of the state. Orange dash line represents the trend line for the northern network.]{\includegraphics[width=7.9cm]{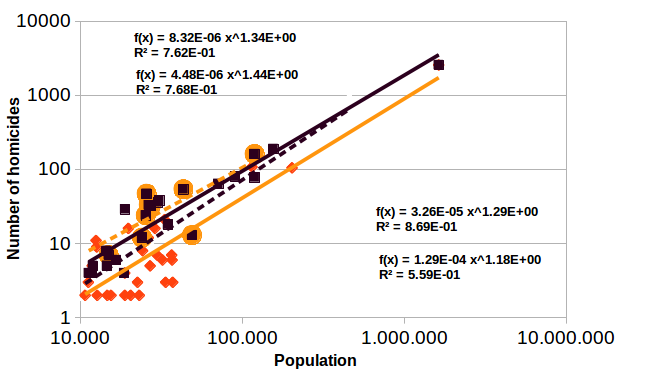}}
\qquad
\subfigure[ESmap][Map of the ES state evidencing the main roads connecting its cities. The shaded area indicates the separation of the two different cities network connected by the main roads of the state.]{\includegraphics[width=4.7cm]{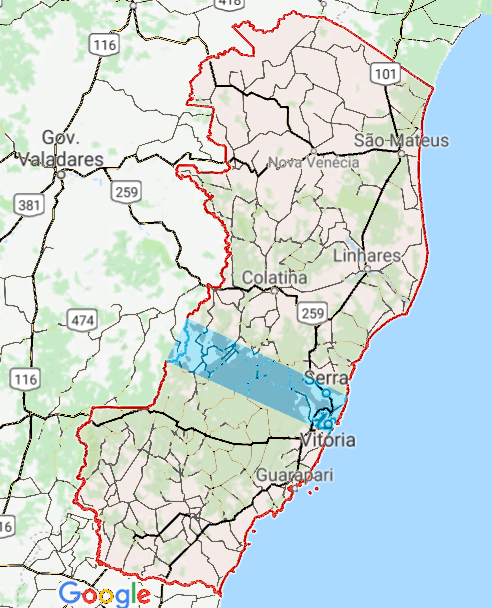}} 
\caption{Scaling behavior of the number of firearms homicides Espirito Santo as analyzed by different methods, and map of the state of Esp\'irito Santo showing the main road crossing the state, modified from a Google Maps image.} \label{ESmapa}
\end{figure}

Maranh\~ao is one of the most dangerous states in Brazil, presenting the largest exponent, $\beta=1.47$. On the other extreme is Santa Catarina, the safest state in the country, for which the slope is $\beta=1.005$, that is, homicides increases practically linearly with population. For reasons that will be apparent soon, we will not investigate the state with the largest slope, Maranh\~ao. Instead, we will analyse in more details the cities in the state of Esp\'irito Santo, which presents the second largest slope found in this analysis, with $\beta=1.44$, and is located in the Southeast region together with Minas Gerais, Rio de Janeiro and S\~ao Paulo. For Esp\'irito Santo, although the small equivalent cities show a relatively small number of homicides as compared to cities in other Brazilian states, the strong increase due to the large slope makes its largest cities to be placed among those with highest homicide rates.

As already mentioned, our assumption is that the exponent $\beta$ obtained for each state varies in a wide range, $1<\beta<1.5$, because for each state one may have cities that do not belong to the same network as the others. Esp\'irito Santo (ES) is a case where this fact may be seen more clearly. For this state highlands in the central region separate the state into two areas, one in the southern region, and one in the central-northern region. There are many paved roads connecting cities in each area, but these two networks are poorly connected one to the other, with the only good connection being the metropolitan region comprising the cities Vit\'oria, Vila Velha, Serra and Cariacica, as can be seen in Fig.~\ref{ESmapa}. These are the four largest cities in the state, located in a semi-circle with radius of about 25 Km from the main city, Vit\'oria, forming a single urban environment. Therefore, geography and bad roads lead to the development of two distinct networks of cities, with only the metropolitan region of Vit\'oria being the single important point of contact. Unfortunately there are no precise information about the flux of people moving from the northern network to the southern network, but there are reasons to believe it is not significant. In fact, Esp\'irito Santo is a small state, with area 46.000 $km^2$ and population of 3.9 million people, as estimated for 2014, most of them, 1.6 million, living in the Vit\'oria metropolitan area, which is the center of socioeconomic activity in the state, as well as an important national touristic spot. ES has a large cost line with beautiful beaches which attract many tourists from Minas Gerais. Aside the metropolitan area, tourists look for cities as S\~ao Mateus, in the north, and Guarapari in the south. While the northern roads connect to Governador Valadares and Te\'ofilo Otoni, the southern roads connect with Belo Horizonte, the lastthree cities being found in the state of Minas Gerais, so one can imagine that the flux of people through each one of the two distinct networks are directed to and fro the metropolitan region or to each of the touristic spots, but scarcely directed to other cities in each network. Therefore we can suppose that cities in the northern part and in the southern part of ES form two distinct networks which characteristics that are distinct. The main roads are directed to the large metropolitan region, and the four cities in this area, Vit\'oria, Vila Velha, Serra and Cariacica, are so connected by an intricate network of roads, avenues and bridges that we can surely consider it as a large region where individual characteristics of each city are merged to form a single urban environment. So we henceforth consider this metropolitan area as one single urban center instead of the four cities separately.

The theoretical framework we are using indicates that each network should present a power-law behavior for the socioeconomic activity. In recent developments of the theory it was argued that the power-law behavior emerges from fractal properties of the city network, so that a small network of cities, if considerably isolated from cities that do not belong to the same network, should present the same characteristics as the larger network it is part of. So, the northern and southern networks found in ES should present characteristics that are similar, and also they should be similar to the national network, which is the main network they are part of. We can expect, therefore, that the exponent $\beta$ for each of the two networks of ES should be closer to the value $\beta=1.15$ found for the country. Of course, the two networks are not really isolated, since they have some connections between them, with the metropolitan region being a common vertex, but also other small connections, even if by terrible roads. Also, they are well connected to cities in other states, as Minas Gerais (MG), Rio de Janeiro (RJ) and Bahia (BA). Therefore, we cannot expect that each of the ES networks will present exactly the value of the national network, but we can expect that their values are closer to $\beta=1.15$ than in the case these two networks are considered as they were a single network.

In Fig.~\ref{ESmapa} we see the results obtained for each of the two networks, as well as the result obtained by considering the whole state as a single network. The values for the coefficient are: $\beta=1.44\,[1.38,1.48]$ in the case the whole state is considered a a single network, and $\beta=1.35\,[1.28,1.37]$ for the northern network and $\beta=1.32\,[1.27,1.36]$ for the southern network, when they are considered separately. Therefore, when the particular characteristics of each network are taken into consideration, we observe the coefficient move towards the expected value, showing the importance of considering isolated networks for the analysis of cities properties according to the power-law theory we are considering in this analysis. We can go further by noticing that in the center-north region another network can be identified with cities well connected among themselves, but poorly connected to other cities, and which instead of being linked directly to the Vit\'oria metropolitan region, it is connected preferentially with S\~ao Mateus, a medium size city in the northern coast of ES. It is interesting to notice that also here the geography plays a role in the isolation of cities, since also here highlands separate this network from the others, although the separation is not as clear as for the southern network. We analyzed the homicide data for the northern network, which does not include the metropolitan area, and it resulted $\beta=1.18\,[1.15,1.20]$, being therefore still closer to the expected value of 1.15. 

\begin{figure}[t]
 \centering
 \includegraphics[scale=0.7]{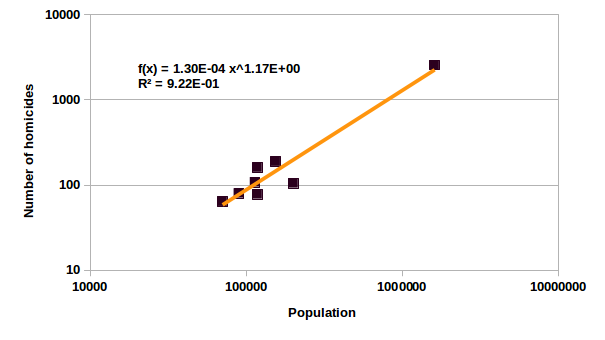} 
 \caption{Scaling of firearms homicides in the main cities in  Esp\'irito Santo state.}  \label{ESmain}
\end{figure}

Another way to restrict the effects external to the network is to confine the analysis to the larger cities which are highly interconnected. In this case, influence of internal influx is diluted due to the city size, and we can expect that the characteristics of their network to be more evident. This is done in the present study for ES state by considering the cities in the state with population larger than $8 \times 10^4$. The results are shown in Fig.~\ref{ESmain}, where we observe that the slope obtained is $\beta=1.17$, in good agreement with the expected value.

It is interesting to compare the increase of firearms homicides in this state with those for equivalent cities in S\~ao Paulo, the second safest state. The rate of crime in Santa Catarina increases at a pace that is somewhat smaller than that for S\~ao Paulo, but for cities with less than one hundred thousand inhabitants, the equivalent cities in S\~ao Paulo are safer than those in Santa Catarina. This is also a surprising result of this analysis, since Santa Catarina is usually considered the safest state in Brazil, no matter the city size. Of course one have to keep in mind that there are large fluctuations around the mean values, and those fluctuations are larger for smaller cities, so the common feeling that Santa Catarina is safer than S\~ao Paulo is reasonable. The state of Santa Catarina has a coefficient that is below the expected value. Here the analysis is more complicated, because this is the safest state in Brazil, the number of cities with 2 or more firearms homicides is small, and the state is formed by a large number of small cities, with the main metropolitan region showing a population smaller than 1 million. In addition, it is an important touristic place in the national level as well as in continental level, with many people from abroad coming during summer time to its beaches, and people from other regions in Brazil traveling to its mountain regions for the winter vacations. In this case, the whole state is not enough to form a single network, when firearms homicides is considered. Our analysis suggests that the influence of tourists in small cities is too large to be neglected, and a network larger than the state is responsible for the overall behavior of cities.

We conclude that when cities are carefully selected according to their more relevant connections, forming a network where the flux of people, services and goods between cities within the network is more important than with cities out of the network, the coefficient $\beta$ tends to values closer to the expected by the theory. In this way the study performed here allows to understand in a deeper way the significance of networks when the power-law theory is applied to study socioeconomic activities. This result reinforce the relation between socioeconomic activity and infrastructure of the network, no matter if the network is comprised in a single city, in a collection of cities or in an entire country. The interesting correspondence between the exponent $\beta=1.15$ for socioeconomic activity and the value $\beta=0.85$~\cite{WestScale} is an evidence that there is a correlation between infrastructure characteristics and socioeconomic activity, and that a more general theory may be necessary to incorporate both aspects in a single theoretical frame.

Of course more definite conclusions could be obtained if the selection of cities in the networks were obtained by considering the real flux of people and vehicles between them. Unfortunately accurate data on flux of people for ES or other Brazilian states are not available. But the infrastructure conditions presented by this particular state allows, with good approximation, to draw evaluate the flux of people between the different networks analyzed here.

\section{Conclusions}

An analysis of firearm homicides crime in Brazilian cities was performed under the light of recent theoretical developments indicating a power-law increase of the number of crimes with the city size. The results confirm the theoretical predictions, showing a power-law behavior with the expected exponent $\beta=1.15$. We also analyzed the behavior of cities for some Brazilian states separately. We observed that the power-law behavior can describe rather well the increase of homicides with the city population even when each state is analyzed separately, with a superlinear increase, that is, with exponent $\beta>1$ for all cases. However, the value of the exponents are spread over a large range between 1 and 1.5. We analyzed in details some cases, in particular for Esp\'isito Santo, for which we obtained $\beta=1.44$, S\~ao paulo, with $\beta=1.15$, and Santa Catarina, with $\beta=1.005$. In the case of Esp\'irito Santo it was possible to identify two or three rather isolated networks of cities that, when considered separately, leads to values of the exponent $\beta$ in better agreement with the expected value.

The results, therefore, confirms the theory and show that the self-similarity observed at the city scale can be observed also in larger scales of networks connecting several cities.
and also allows for a deeper understanding of the importance of considering isolated networks for the analysis. Although the study is not conclusive, due to the fact that only firearms homicides is considered, this study indicates that the idea that cities develop according to fractal networks is sound, and should be better investigate using other aspects of socioeconomic activity, preferably in connection with infrastructure networks, such as traffic flux between cities, in order to get a complete understanding of the network structure.

{\bf Acknowledgments:} This work was supported by the Conselho Nacional de Desenvolvimento
Científico e Tecnológico (CNPq-Brazil).


\vspace{0.5cm}

\begin{thebibliography}{1}
\bibitem{Mumford} L. Mumford, ``The City in History: Its Origins, Its Transformations, and Its Prospects'', 1961, New York, Brace \& World.
\bibitem{WestCityNature} L.M.A. Bettencourt and G. B. West, ``A Unified Theory of Urban Living'', Nature 467 (2010) 912.
\bibitem{Batty} M. Batty, ``The New Science of Cities'', MIT Press, Cambridge, MA, 2014. 
\bibitem{Netto} Vinicius M. Netto, Edgardo Brigatti, Jo\~ao Meirelles, Fabiano L. Ribeiro, Bruno Pace, Caio Cacholas and Patricia Sanches, ``Cities, from Information to Interaction'', Entropy 20 (2018) 834.
\bibitem{Bettencourt1} L.M.A. Bettencourt, J. Lobo, D. Strumsky and G.B. West, 2010 Urban Scaling and Its Deviations:
Revealing the Structure of Wealth, Innovation and Crime across Cities PLoS ONE 5 e13541
\bibitem{Bettencourt2} L.M.A. Bettencourt and G. B. West, 2011 Bigger cities do more with less Scientfic American 305
52-53
\bibitem{Bettencourt3} L.M.A. Bettencourt, J. Lobo, D. Helbing, K F Chnert and G.B. West 2007 Growth, innovation,
scaling, and the pace of life in cities Proc. Natl. Acad. Sci. USA 104 7301-7306
\bibitem{Batty2} M. Batty 2008 ``The size, scale, and shape of cities'' Science 319 769-771.
\bibitem{WestScale} G. B. West, ``Scale: The Universal Laws of Life, Grouth, and Death in Organisms, Cities and Companies'', 2017, Penguin Book, New York.
\bibitem{Chowell} Chowell G, Hyman J M, Eubank S and Castillo-Chavez C 2003 Scaling laws for the movement
of people between locations in a large city Phys. Rev. E 68 066102.
\bibitem{BattyFractalCities} M. Batty and P. Longley, ``Fractal Cities: A Geometry of Form and Function'', Academic Press, Cambridge, MA, 1994.
\bibitem{Schlapfer} M. Schl\"apfer et al., ``The Scaling of Human Interactions with City Size'', Journal of the Royal Society Interface 11 (2014) 20130789.
\bibitem{crimes1} M. Oliveira, C. Bastos Filho and R. Menezes, (2017) ``The scaling of crime concentrtion in cities'', PLoS ONE 12 (8): e0183110.
\bibitem{crimes2} Ribeiro HV, Hanley QS, Lewis D (2018) ``Unveiling relationships between crime and property in England and Wales via density scale-adjusted metrics and network tools''. PLoS ONE 13(2):
e0192931.
\bibitem{Mello} M. H. P. Mello-Jorge, ``Como Morrem Nossos Jovens''. In: ``CNPD. Jovens Acontecendo na Trilha das Pol\'iticas P\'ublicas''. Bras\'ilia, 1998.
\bibitem{Ramos} R. de Souza et. al., ``Qualidade da informa\c{c}\~ao sobre viol\^encia: um caminho para a constru\c{c}\~ao da cidadania'',  In: ``INFORMARE - Cadernos do  Programa  de  P\'os-Gradua\c{c}\~ao  em  Ci\^encias  da  Informa\c{c}\~ao'',  Rio  de Janeiro, v. 2, n. 1, jan./jun, 1996.
\bibitem{MV} J. J. Waiselfiz, ``Mapa da Violecia 2016'', https://www.mapadaviolencia.org.br. 
\bibitem{datasus} http://www2.datasus.gov.br.
\bibitem{BattyComplexity} M. Batty, ``Cities and Complexity'', 2014, MIT Press, Cambridge MA.
\bibitem{Bettencourt4} Bettencourt L M A, ``The Origins of Scaling in Cities'', Science 340 (2013) 1438-1441.
\bibitem{Youn} H. Youn H, L.M.A. Bettencourt,
J. Lobo, D. Strumsky, H. Samaniego, G.B. West. 
``Scaling and universality in urban econ-
omic diversification'', J. R. Soc. Interface 13 (2016) 20150937.
\bibitem{Bettencourt-India} Sahasranaman A,Bettencourt LMA. 2019 Urban geography andscaling of contemporary Indian cities.J.  R.  Soc.Interface16: 20180758.
%
\end{thebibliography}
\end{document}